\documentclass[a4paper,10pt,twocolumn]{article}
\pdfoutput=1
\usepackage[a4paper, total={18cm, 26cm}]{geometry}

\usepackage{cite}
\usepackage{amsmath,amssymb,amsfonts}
\usepackage{graphicx}
\usepackage{textcomp}

\usepackage{acronym}
\usepackage{subfig}
\usepackage[per-mode=symbol]{siunitx}
\usepackage[hidelinks]{hyperref}
\usepackage{booktabs}
\usepackage{xspace}
\usepackage{algorithm} 
\usepackage{algpseudocode}
\usepackage[font={small}]{caption}
\usepackage{color}
\newcommand{\red}[1]{{\color{red}#1}}

\usepackage{hyphenat}
\usepackage{tabularx}
\usepackage{multirow}

\usepackage[capitalise]{cleveref}
\crefname{section}{Sect.}{Sect.}
\crefname{figure}{Fig.}{Fig.}
\crefname{table}{Tab.}{Tab.}
\crefname{equation}{Eq.}{Eq.}
\def\figname{\csname cref@figure@name\endcsname\xspace}
\def\tabname{\csname cref@table@name\endcsname\xspace}
\def\secname{\csname cref@section@name\endcsname\xspace}
\def\Secname{\csname Cref@section@name\endcsname\xspace}
\def\secsname{\csname cref@section@name@plural\endcsname\xspace}
\def\Secsname{\csname Cref@section@name@plural\endcsname\xspace}
\def\eqname{\csname cref@equation@name\endcsname\xspace}
\def\eqpname{\csname cref@equation@name@plural\endcsname\xspace}

\acrodef{3GPP}{3rd Generation Partnership Project}
\acrodef{4G}{4$^{\text{th}}$ Generation}
\acrodef{5G}{5$^{\text{th}}$ Generation}
\acrodef{5GV2X}[5G-V2X]{5$^{\text{th}}$ Generation Cellular V2X}
\acrodef{ACC}{Adaptive Cruise Control}
\acrodef{ACI}{Adjacent Channel Interference}
\acrodef{AoI}{Age of Information}
\acrodef{API}{Application Programming Interface}
\acrodef{AWGN}{Additive White Gaussian Noise}
\acrodef{BER}{Bit Error Probability}
\acrodef{BSM}{Basic Safety Message}
\acrodef{CACC}{Cooperative Adaptive Cruise Control}
\acrodef{CAM}{Cooperative Awareness Message}
\acrodef{CAV}{Cooperative Automated Vehicle}
\acrodef{CFD}{Computational Fluid Dynamics}
\acrodef{CSH}{Constant Space Headway}
\acrodef{CTH}{Constant Time Headway}
\acrodef{CSMA}{Carrier-Sense Multiple Access}
\acrodef{CV2X}[C-V2X]{Cellular V2X}
\acrodef{D2D}{Device-to-Device}
\acrodef{DENM}{Decentralized Environmental Message}
\acrodef{DES}{Discrete Event Simulation}
\acrodef{DSRC}{Direct Short Range Communications}
\acrodef{EIDM}{Enhanced Intelligent Driver Model}
\acrodef{eNB}{eNodeB}
\acrodef{egnb}[e(g)NB]{e(g)NodeB}
\acrodef{FCC}{Federal Communications Commission}
\acrodef{FDMA}{Frequency-Division Multiple Access}
\acrodef{FOT}{Field Operational Test}
\acrodef{gNB}{gNodeB}
\acrodef{GUI}{Graphical User Interface}
\acrodef{HARQ}{Hybrid Automatic Repeat reQuest}
\acrodef{HBEFA}{HandBook Emission FActors for road transport}
\acrodef{IC}{In Coverage}
\acrodef{IDM}{Intelligent Driver Model}
\acrodef{IVC}{Inter Vehicle Communication}
\acrodef{LTE}{Long Term Evolution}
\acrodef{LTEV2X}[LTE-V2X]{Cellular V2X}
\acrodef{MAC}{Medium Access Control}
\acrodef{MEC}{Mobile Edge Computing}
\acrodef{MPC}{Model Predictive Control}
\acrodef{NR}{New Radio}
\acrodef{OoC}{Out of Coverage}
\acrodef{PC5}{PC5}
\acrodef{PDR}{Packet Delivery Ratio}
\acrodef{RADCOM}{RADar-based COMmunication}
\acrodef{RB}{Resource Block}
\acrodef{RSU}{Road Side Unit}
\acrodef{SAE}{Society of Automotive Engineering}
\acrodef{SBSPS}{Sensing-Based Semi-Persistent Scheduling}
\acrodef{SC-FDMA}{Single Carrier-Frequency Division Multiple Access}
\acrodef{SINR}{Signal to Interference plus Noise Ratio}
\acrodef{SIR}{Signal to Interference Ratio}
\acrodef{SNR}{Signal to Noise Ratio}
\acrodef{SRSSI}[S-RSSI]{Side Link Received Signal Strength Indicator}
\acrodef{SUMO}{Simulation of Urban MObility}
\acrodef{TA}{Traffic Authority}
\acrodef{TB}{Transport Block}
\acrodef{TDMA}{Time-Division Multiple Access}
\acrodef{TraCI}{Traffic Control Interface}
\acrodef{UE}{User Equipment}
\acrodef{V2V}{Vehicle to Vehicle communication}
\acrodef{V2X}{Vehicle to Everything}
\acrodef{VRU}{Vulnerable Road Users}
\acrodef{WAVE}{Wireless Access in Vehicular Environments}
\acrodef{ZOH}{Zero Order Hold}

\newcommand{\plexe}{\textsc{Plexe}\xspace}
\newcommand{\ploeg}{\textsc{Ploeg}\xspace}
\newcommand{\pathc}{\textsc{PATH}\xspace}
\newcommand{\gsbl}{\textsc{GSBL}\xspace}

\newcommand{\Cruise}{\textsc{Cruise}}
\newcommand{\Override}{\textsc{Override}}

\newcommand{\ca}{\ensuremath{\texttt{A}}\xspace}
\newcommand{\cpa}{\ensuremath{\texttt{P}}\xspace}
\newcommand{\cpl}{\ensuremath{\texttt{L}}\xspace}
\newcommand{\cgs}{\ensuremath{\texttt{G}}\xspace}

\newcommand{\noc}{\ensuremath{N}\xspace}
\newcommand{\tdns}{\ensuremath{D_v}\xspace}
\newcommand{\nlane}{\ensuremath{M_{L}}\xspace}
\newcommand{\thrN}{\ensuremath{\Theta_{\text{\footnotesize{N}}}}\xspace}
\newcommand{\thrE}{\ensuremath{\Theta_{\text{\footnotesize{E}}}}\xspace}
\newcommand{\thrS}{\ensuremath{\Theta_{\text{\footnotesize{S}}}}\xspace}
\newcommand{\thrW}{\ensuremath{\Theta_{\text{\footnotesize{W}}}}\xspace}

\newcommand{\udes}{\ensuremath{\dot{u}}\xspace}
\newcommand{\ades}{\ensuremath{u}\xspace}

\newcommand{\ddes}{\ensuremath{d_d}\xspace}

\newcommand{\ra}{\ensuremath{a}\xspace}
\newcommand{\rv}{\ensuremath{v}\xspace}
\newcommand{\rp}{\ensuremath{x}\xspace}
\newcommand{\thead}{\ensuremath{H}\xspace}
\newcommand{\cfv}{\ensuremath{\xi}\xspace}

\newcommand{\vrefg}{\ensuremath{v_r}\xspace}

\newcommand{\ccc}{\ensuremath{\mathbf{C}}\xspace}

\DeclareSIUnit \kmh {\kilo\meter\per\hour}
\DeclareSIUnit \mps {\meter\per\second}
\DeclareSIUnit \mpsq {\meter\per\second\squared}
\DeclareSIUnit \mpsc {\meter\per\cubic\second}

\def\BibTeX{{\rm B\kern-.05em{\sc i\kern-.025em b}\kern-.08em
    T\kern-.1667em\lower.7ex\hbox{E}\kern-.125emX}}

\begin{document}
\sloppy
\title{\vspace{-0.5cm}\bfseries Heterogeneous CACC Coexistence: Simulation, Analysis, and Modeling}

\author{Lorenzo Ghiro$^1$, Marco Franceschini$^2$, Renato Lo Cigno$^1$, Michele Segata$^3$\\[3pt]
\textit{$^1$Dep. of Information Engineering, University of Brescia, Italy}\\
\textit{$^2$CHECK24 GmbH, Munich, Germany}\\
\textit{$^3$Dep. of Information Engineering and Computer Science, University of Trento, Italy}\\[3pt]
\thanks{This work was partially supported by the Italian Ministry for Universities and Research (MUR), National Recovery and Resilience Plan (NRRP), at the University of Brescia with projects ``Sustainable Mobility Center (MOST)'' Spoke N$^o$ 7, ``CCAM, Connected networks and Smart Infrastructures'' -- CUP D83C22000690001, and PROSDS, part of RESTART, Spoke 4, Structural Project SUPER -- CUP C89J24000270004; at the University of Trento with NRRP PRIN ``Self-optimizing Networked Edge Control for Cooperative Vehicle Autonomy (SELF4COOP)" -- CUP E53D23000920001. 
Views and opinions expressed are those of the authors only and do not necessarily reflect those of the funding institutions nor the granting authority can be held responsible for them.}
\vspace{-0.5cm}}

\date{~}

\maketitle

\begin{abstract}%
The design of \ac{CACC} algorithms for vehicle platooning has been extensively investigated, leading to a wide range of approaches with different requirements and performance. 
Most existing studies evaluate these algorithms under the assumption of homogeneous platoons, i.e., when all platoon members adopt the same \ac{CACC}. 
However, market competition is likely to result in vehicles from different manufacturers implementing distinct \acp{CACC}. 
This raises fundamental questions about whether heterogeneous vehicles can safely cooperate within a platoon and what performance can be achieved. 
To date, these questions have received little attention, as heterogeneous platoons are difficult to model and analyze.

In this work, we introduce the concept of \emph{mixed platoons}, i.e., platoons made of vehicles running heterogeneous \acp{CACC}, and we study their performance through simulation-based experiments.
We consider mixtures of three well-established \acp{CACC} from the literature. 
In the first part of the paper we study a single mixed platoon in isolation to understand the microscopic effects on safety:
we evaluate the performance of various CACC-mixtures across speed change and emergency braking scenarios.
In the second part, we examine a high-density ring-road scenario to assess macroscopic impacts on safety, comfort, and traffic throughput,
especially comparing throughput results with those obtained from vehicles controlled by a standard \ac{ACC} or by human-drivers. 

Our findings highlight that some combinations of \acp{CACC} can operate robustly and safely, while others exhibit critical limitations in safety, comfort, or efficiency.
These results emphasize the need for careful system design and the development of theoretical frameworks for modeling heterogeneous platoons.
\end{abstract}

\acresetall
\section{Introduction}
\label{sec:introduction}

\acp{CACC}, the foundation of vehicle platooning, are one of the cooperative driving applications that has attracted significant attention from both academia and industry. Despite this interest and the clear benefits it promises, large-scale deployment is still lagging behind. Experimental studies remain limited due to high costs and typically address only small platoons in which all vehicles implement the same control algorithm. 

Heterogeneity in vehicle type, such as differences in mass or performance, has been considered to some extent in the literature, but it is usually simplified by constraining all vehicles to operate at the level of the least capable one; for instance, a sports car can be forced to behave like a heavy truck. In contrast, heterogeneity in terms of the control algorithms implemented on board, where vehicles in the same platoon rely on different \acp{CACC}, has received even less attention and remains largely unexplored.
After several years without major developments, \ac{SAE} released its first recommendation for \ac{CACC} in October 2023\footnote{See \url{www.sae.org/standards/content/j2945/6/}, last visited September 2025.}. 
This standard explicitly states that it targets systems operating under driver responsibility and supervision, and that it \textit{uses a time-gap control strategy similar to ACC}. 

Although a standardized \ac{CACC} algorithm may seem a reasonable solution, technical and commercial factors are likely to result in diverse implementations. 
This makes it important to ask what happens when vehicles equipped with different \ac{CACC} algorithms and controllers meet on the road, and how they interact both with vehicles using conventional \ac{ACC} or with the enhanced version under study by \ac{SAE}. 

This situation raises several research questions. 
From an academic perspective, it is essential to investigate whether a theoretical framework can be developed to formally characterize the properties of such heterogeneous formations. 
Initial efforts in this direction exist, particularly in the field of consensus under changing control topologies, initiated by the seminal work of~\cite{ren2005consensus}.
This line of research has since expanded in various directions~\cite{zhu2015necessary, liu2021consensus} and has also been applied to networked vehicular control~\cite{santini2017consensusbased, li2019consensusbased, santini2019platooning}. 

From a practical perspective, however, only limited attempts have been made to evaluate what occurs when \acp{CAV} equipped with different \acp{CACC} interact in real traffic. 
Key questions remain unanswered: can such vehicles safely coexist, should they fall back to a common control algorithm if that were feasible, and will overall road performance --in terms of throughput-- be impaired or instead enhanced?

This paper builds on our preliminary work in~\cite{segata2021progressive} and provides new insights into the possibilities, risks, and potential of mixing different longitudinal controllers in flows of connected and cooperative vehicles. As a final contribution, in \cref{sec:matrix} we propose a potential approach to formally model heterogeneous platoons, thereby paving the way for future research on this topic.

\section{Related Work}
\label{sec:related}

The literature on longitudinal platoon controllers is extensive and relies on different theoretical frameworks. Among the most widely used controllers are \ploeg~\cite{ploeg2011design} and \pathc~\cite{rajamani2012vehicle}, both based on traditional control theory but targeting different objectives. \ploeg aims to replicate the behavior of standard \ac{ACC} while improving performance and string stability through a \ac{CTH}, whereas \pathc maintains \ac{CSH} and relies on information from both the preceding vehicle and the platoon leader. A third proposal, \gsbl~\cite{giordano2017joint,giordano2019joint}, focuses on robustness to information loss and the ability to follow an external reference speed $\vrefg(t)$, exploiting information from both front and rear neighbors in a spring-damper fashion. We adopt these three controllers in our study and provide details in \cref{ss:controllers}.  

Other works extend traditional control theory with different perspectives. 
For example,~\cite{besselink2017string} analyzes controllers in the spatial rather than the temporal domain, noting that platoon speed profiles are often linked to road position rather than to headway time.
Authors of~\cite{qin2022nonlinear} propose a nonlinear vehicle-following model inspired by human driving, which resembles \ploeg as it can also exploit communications, but has not been implemented in simulation.

Consensus theory offers an alternative approach where the control topology is part of the design space rather than a constraint. 
Stability is usually studied via Lyapunov-Razumikhin methods, enabling analysis under arbitrary topologies and delays~\cite{santini2017consensusbased, bernardo2014distributed}, and also maneuvers modeled as changes in control topology~\cite{santini2019platooning}. This line of research has grown considerably~\cite{li2019consensusbased, yang2019consensus, li2018nonlinear, zheng2023development}, and some works implicitly touch on heterogeneity, for instance by mixing \ac{CTH} and \ac{CSH} policies within a platoon~\cite{zheng2023development}.  

Only a few studies explicitly address heterogeneous controllers. Early work~\cite{gong2024cav-adv} models \ac{CACC} and human-driven vehicles within a unified framework and analyzes stability via $H_{\infty}$ methods, though only in simplified scenarios. Other efforts consider issues related to delays~\cite{flores2018fractionalorder, wang2020optimal} or vehicle heterogeneity~\cite{liu2024selforg}, but not the coexistence of multiple control laws. Similarly, approaches based on graceful degradation~\cite{liu2023graceful}, sub-platoon formation~\cite{zhang2025selforganized}, or learning-based delay compensation~\cite{tian2021lstm, zhang2025lstm} relate to robustness but not to mixing different controllers.  

Closer to our focus,~\cite{qin2019rearend, segata2022multi-technology} analyze fallback procedures from \ac{CACC} to \ac{ACC} in case of communication failures or attacks~\cite{wons25misbe}, considering \ploeg, \pathc, and standard \ac{ACC}. In these works, however, different controllers are used sequentially during transitions rather than stably coexisting. More explicit attempts at heterogeneous platoons include~\cite{zhou2019stabilizing}, which derives stability conditions for platoons mixing \ac{CACC} and human-driven vehicles, and~\cite{qin2020string}, which approximates stability under varying penetration rates of \pathc. Still, these analyses remain limited in scale or scope.  

In summary, most literature that refers to heterogeneity focuses on vehicle characteristics rather than control laws~\cite{viaderomonasterio2024robust,liu2023distributed,liu2024learning,saito2024network}. Only a handful of studies examine the coexistence of different controllers, and even then under restrictive assumptions. This motivates our work, which does not propose a new \ac{CACC}, but instead investigates the impact of mixing existing controllers on both platoon-level dynamics and overall traffic performance. For this purpose, we select \ploeg, \pathc, and \gsbl as representative controllers: \ploeg as an improved \ac{ACC}-like scheme, \pathc for its efficiency-oriented constant spacing, and \gsbl for its robustness to communication impairments. Together, these capture diverse design goals and control topologies, making them a suitable basis for studying heterogeneous platoons. 

Robustness and resilience aspects are studied elsewhere~\cite{giordano2019joint, ploeg2015graceful, wu2019cooperative}, while large-scale stability analysis of multiple parallel platoons has been addressed in~\cite{li2019consensusbased, weiwei2025multiplatoon}, but always under the assumption of homogeneous controllers.

\section{CACCs Considered}
\label{sec:modeling}

\begin{table}
\centering
\small
\begin{tabular}{p{1.25cm} c p{3cm} p{1.2cm}} 
\toprule
\textbf{Alg.}      & V2X   & Topology & Spacing \\     
                         & Y/N   &     &      \\     
\midrule                               
\ac{ACC}~\cite{rajamani2012vehicle} & N & Predecessor &  \ac{CTH} \\
\midrule
\ploeg~\cite{ploeg2011design}       & Y & Predecessor &  \ac{CSH}\\
\midrule                               
\pathc~\cite{rajamani2012vehicle}   & Y & Leader, Predecessor &  \ac{CTH}\\
\midrule
\gsbl~\cite{giordano2019joint}      & Y & Leader, Predecessor, Follower & \ac{CSH} \\
\bottomrule 
\end{tabular}
\caption{Summary of the key features of the longitudinal control algorithms considered.}
\label{tab:cacc}
\vspace{-3mm}
\end{table}
 
This work does not focus on any specific communication technology such as 802.11p or \ac{CV2X}; rather, it aims to understand whether and how heterogeneous \ac{CACC} capabilities can be progressively introduced without destabilizing existing traffic, while potentially improving overall performance and road usage. For this reason, we assume that vehicles communicate reliably with an update frequency of at least \SI{10}{\hertz}, exchanging all the parameters required by the considered control algorithms.  

We consider a highway scenario where all vehicles are either driven by humans following the \ac{EIDM} model~\cite{salles2020extending} or are \ac{ACC} capable, and investigate the progressive introduction of \ac{CACC} capabilities, either all with the same controller or with three different controllers mixed together randomly, as if building platoons on-the-fly with different technologies when vehicles encounter each other on the road. 
The actuation lag of each vehicle is modeled as a first-order low-pass filter with a pole at \SI{0.5}{\second}, a widely accepted reference model~\cite{rajamani2012vehicle}.  
\Cref{tab:cacc} summarizes the main features of the controllers employed, which are described in detail in the remainder of this section to ensure the paper is self-contained. 

\subsection{Controllers Included in the Study}
\label{ss:controllers}

The \acf{ACC} we adopt is the classical one defined in \cite[Chapter~6]{rajamani2012vehicle}.
Its \ac{CTH} control law is defined as
\begin{align}
    \ades_i & = - \frac{1}{\thead} \left(\dot{\varepsilon}_i + \lambda\delta_i\right) \label{eq:acc}\\
    \delta_i & = \rp_i - \rp_{i-1} + l_{i-1} + \thead\rv_i; \qquad    \dot{\varepsilon}_i = \rv_i - \rv_{i-1} \nonumber
\end{align}
where $l_i$, $\ades_i$, $\rp_i$, $\rv_i$ are the length, control input (desired acceleration), position, and speed of the $i$-th vehicle, respectively, and $\lambda$ and \thead are the controller parameters that define the desired headway time.
$\dot{\varepsilon}_i$ is the speed difference from the vehicle in front, and $\delta_i$ is the bumper-to-bumper distance error from the same vehicle. 
We consider $\thead=\SI{1.2}{\second}$, a value ensuring string stability and comfort.
In the rest of the paper we also call it the \ca controller. 

As \ac{CACC} controllers we consider \ploeg \cite{ploeg2011design}, \pathc \cite{rajamani2012vehicle} and \gsbl \cite{giordano2019joint} for the reasons discussed in \cref{sec:related}. 
\ploeg controller,  \cpl in the rest of the paper, is also based on \ac{CTH} policy, thus mimicking an \ac{ACC} behavior and using the same control topology, but the string stability and performance are improved thanks to \ac{V2X} communications that distribute the control input $\ades_{i-1}$ for the benefit of vehicle $i$, thus reducing the reaction time of the following vehicle because it discounts the actuation lag of the power train. 
\ploeg's control law is defined as:
\begin{align}
\udes_i & = \frac{1}{\thead}(-\ades_i + k_p (\rp_{i-1} - \rp_i - l_{i-1} - \thead\rv_i) \nonumber \\
            & + k_d (\rv_{i-1} - \rv_i - \thead\ra_i) + \ades_{i-1})
\label{eq:ploeg}
\end{align}
with $k_p$ and $k_d$ parameters controlling how much the controller weights distance and speed errors.
We use the values proposed in the original paper which correspond to $\thead=\SI{0.5}{\second}$.

\pathc controller, denoted as \cpa, is instead defined as:
\begin{align}
\ades_i & = \alpha_1 \ades_{i-1} + \alpha_2 \ades_0 + \alpha_3 (\rv_i - \rv_{i-1}) \nonumber \\ 
            & + \alpha_4 (\rv_i - \rv_0) + \alpha_5 (\rp_i - \rp_{i-1} + l_{i-1} + \ddes) 
\label{eq:path}
\end{align}
where
\vspace{-0.5em}
\begin{align}
    \alpha_1 & = 1 - C_1; \quad \alpha_2 = C_1; \quad  \alpha_5 = -\omega_n^2 \nonumber \\
    \alpha_3 & = -\left(2\xi - C_1\left(\xi + \sqrt{\xi^2 - 1}\right)\right)\omega_n \nonumber \\
    \alpha_4 & = -C_1\left(\xi + \sqrt{\xi^2 - 1}\right)\omega_n \nonumber
\end{align}
with parameters $C_1$, $\xi$, and $\omega_n$ controlling the apportioning of acceleration between
leading and preceding vehicles, damping ratio, and bandwidth, respectively;
\ddes instead is the desired inter-vehicle distance and the remaining variables keep their already defined meaning.
\cpa targets a constant inter-vehicle distance independent from the platoon speed, so it is quite different from \ca and \cpl. 

\gsbl controller, identified as \cgs, is defined by the following equation set:
for vehicle $0$ (the front vehicle),
\begin{align}\label{eq:abs1}
u_0 =  - k(x_{0} -x_{1}-d) - h(v_{0} -v_{1}) -r(v_{0}-\vrefg) + \delta_1 
\end{align}
for vehicles $i=1,\dots N-2$,
\begin{align}\label{eq:absi}
u_i &=  -k(x_{i} -x_{i+1}-d) -k(x_{i} -x_{i-1}+d) \nonumber\\
&\quad -h(v_{i} -v_{i+1}) -h(v_{i} -v_{i-1})-r(v_{i}-\vrefg) + \delta_i 
\end{align}
and, for vehicle $N-1$ (the last one),
\begin{align}\label{eq:absN}
u_{N-1} &= -k(x_{N-1} -x_{N-2}+d) -h(v_{N-1} -v_{N-2}) \nonumber\\
&\quad -r(v_{N-1}-\vrefg) + \delta_{N-1}
\end{align}
where $k$, $h$, and $r$ are parameters of the controller, while $\delta_i$ is a disturbance derived from communication losses and impairments and \vrefg is the (time-varying) reference speed for the entire platoon. 

\section{Mixed Platoons Formation}
\label{s:mixed-met}

For clarity, we define a \textit{platoon} as a set of \noc vehicles driving with communication-based cooperation. 
Vehicles in a platoon $\cal{P}$ are numbered $V_0, \ldots, V_{\noc-1}$, with $V_0$ being the first vehicle, which follows an independent speed profile associated with standard \ac{ACC}. 
The following vehicles may implement any mix of the \pathc, \ploeg, or \gsbl controllers.

With this definition, a generic platoon $\cal{P}$ is identified by the sequence of controllers implemented by its $N_{p}$ vehicles, with a ``$-$'' in position $0$ if $V_0$ follows an independent driving profile. 
For instance, the sequence $\{-, \cpa, \cpl, \cpl \}$ identifies a four-vehicle platoon where $V_0$ is independent, $V_1$ implements \pathc, and $V_2$ and $V_3$ implement \ploeg. 
Similarly, $\{\cgs, \cgs, \cgs, \cpl \}$ describes a platoons with 4 members, the first three adopting \gsbl while the last one is a \ploeg vehicle. 
The choice of control algorithm for each vehicle directly determines the communication pattern, or \textit{control topology}, since each controller relies on information from a specific subset of vehicles.  

\begin{figure}
\centering
\includegraphics[width=0.9\columnwidth]{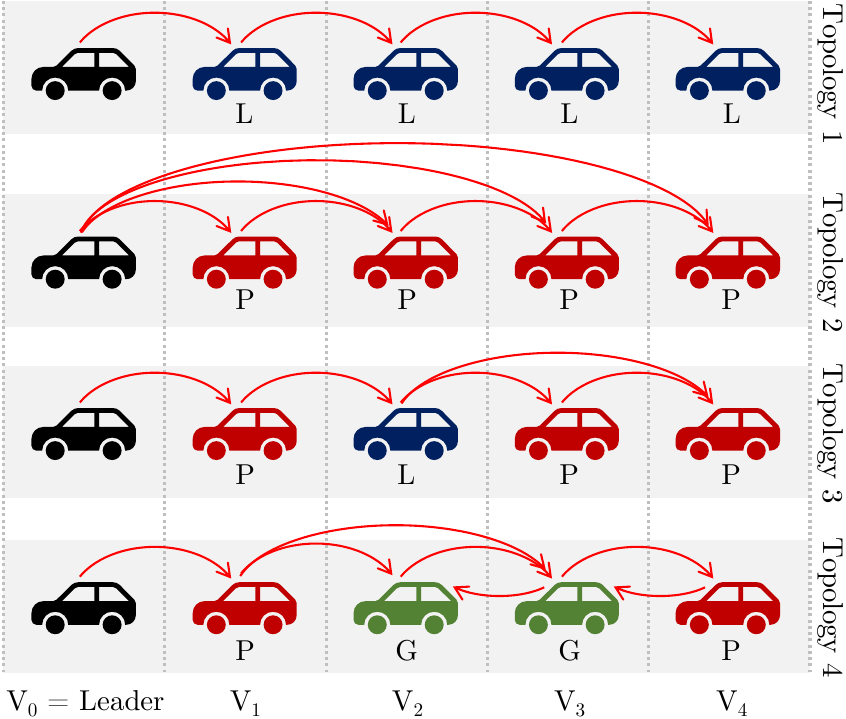}
\caption[width=.9\columnwidth]{Examples of control topology in platoons of five vehicles: Topology 1 $\{-, \cpl, \cpl, \cpl, \cpl \}$; Topology 2 $\{-, \cpa, \cpa, \cpa, \cpa \}$; Topology 3 $\{-, \cpa, \cpl, \cpa, \cpa \}$; Topology 4 $\{-, \cpa, \cgs, \cgs, \cpa \}$.}
\label{fig:topo}
\end{figure}

\Cref{fig:topo} shows four examples of platoons with five vehicles each, with $V_0$ always following its own independent \ac{ACC} profile. 
The first platoon is homogeneous and composed of \cpl controllers, yielding the well-known predecessor-following topology. 
The second, also homogeneous, contains only \cpa controllers, where vehicles also receive information directly from the leader. 
The bottom two illustrate mixed platoons. 
In Topology 3, $V_1$, $V_3$, and $V_4$ implement \cpa, while $V_2$ implements \cpl. 
In Topology 4, $V_1$ and $V_4$ implement \cpa, while $V_2$ and $V_3$ implement \cgs. 
As these examples show, the control topology changes substantially depending on the mix of algorithms. 
In fact, the very notion of a platoon \textit{leader} becomes less clear in mixed settings. 
As discussed in \cref{ssec:mixedleaders}, each vehicle $V_i$ elects as its \textit{egoLeader} the first vehicle ahead that uses a control algorithm different from $V_i$.

\subsection{Leader elections in mixed platoons}
\label{ssec:mixedleaders}

Consider Topology 3 in \cref{fig:topo}. 
If $V_3$ and $V_4$ rely on the information broadcast by $V_0$, the platoon may become destabilized, since the behavior of $V_2$ (\cpl) is not consistent with \cpa.  
If instead they elect $V_2$ as their leader, they can successfully follow it, as the leader's behavior is generally allowed to differ from that of the followers' \ac{CACC}. 
This aspect is even more critical in the case of \gsbl, where a vehicle also incorporates information from the rear neighbor.  
Topology 4 in \cref{fig:topo} illustrates that $V_2$ and $V_3$ elect $V_1$ as their leader, while $V_4$ elects $V_3$.  
It is important to stress that this does not imply splitting the platoon into multiple sperated platoons: 
communication continues to bind all vehicles together, even if through indirect references to intermediate leaders.  

\begin{figure}
    \centering
    \includegraphics[width=\columnwidth]{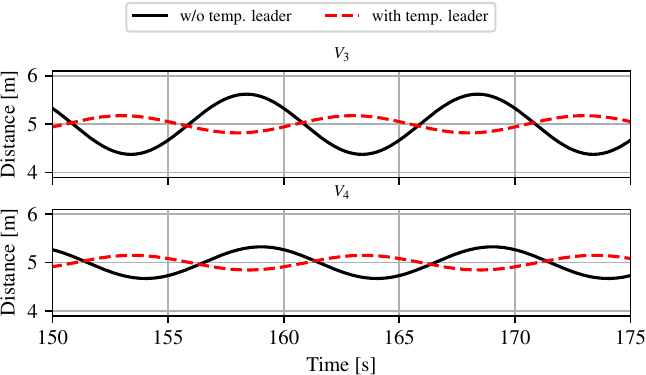}
    \caption[width=.9\columnwidth]{Comparison of the inter-vehicle distance dynamics for $V_3$ and $V_4$ in Topology 3 of \cref{fig:topo}, with (dashed red) and without (solid black) electing $V_2$ as their leader in the \ac{CACC} algorithm.}
    \label{fig:nosubplatoon}
\end{figure}

\Cref{fig:nosubplatoon} highlights the effect for a $\{-, \cpa, \cpl, \cpa, \cpa \}$ platoon by plotting the inter-vehicle distances $V_2 \leftarrow V_3$ and $V_3 \leftarrow V_4$ when $V_0$ follows a sinusoidal speed profile with period \SI{10}{\second}. 
If $V_3$ and $V_4$ elect $V_2$ as their leader, the control topology corresponds to Topology 3 in \cref{fig:topo}; if not, they continue to use $V_0$ as their reference. 
The top plot in the figure refers to $V_2 \leftarrow V_3$, while the bottom one refers to $V_3 \leftarrow V_4$. 
It is evident that the oscillations of $V_3$ are larger when $V_0$ is considered the leader, and the same holds for $V_4$, although with smaller amplitude. 
In both cases, the oscillations are in counter-phase due to the different behavior of \ploeg compared with \pathc.  

Based on these observations, in the remainder of the paper we assume that each vehicle $V_i$ elects its \textit{egoLeader} as $V_j$, according to \cref{eq:egoleader}:
\begin{equation}
V_j: j = \max( \{j~|~j \in \{i-1,\ldots,0\} \wedge \textrm{CTR}_j \neq \textrm{CTR}_i\} )
\label{eq:egoleader}
\end{equation}
where CTR$_j$ denotes the controller of vehicle $j$.  

This heuristic is only one possible method for establishing an \textit{egoLeader} election mechanism, and we believe it deserves further investigation in future studies. Nevertheless, it represents a reasonable assumption for exploring how \ac{CACC}-enabled vehicles may spontaneously form platoons in the presence of heterogeneous control algorithms.

\subsection{Heuristic for safely mixing \gsbl}

It is evident, even without dedicated experiments, that some \acp{CACC} cannot be straightforwardly combined with others. 
This is the case for \gsbl. 
The original article~\cite{giordano2019joint} first introduces the control law for standard cruising and later describes the modifications required to handle external events, since the baseline cruising algorithm alone cannot react promptly. 
Such events include emergency braking, keeping a safe distance from a slower vehicle, or adapting to infrastructure-based speed advisories.  

\Secsname~6.2~and~6.3 in~\cite{giordano2019joint} define an \textit{override mode} that dynamically adjusts controller parameters so that the ego vehicle reacts more aggressively under external disturbances, thereby reducing safety risks. 
However, this mechanism assumes that all vehicles use the same control law.
In particular, \Secname~6.3 describes how to maintain a safe distance from a slower leader by assuming that the leader runs \ac{ACC} in parallel with \gsbl.
Switching between cruising and override modes is then defined to guarantee a headway at least equal to that of \ac{ACC}. 
In a mixed setting, this assumption is problematic: the first \gsbl vehicle would be forced to keep a large gap from the vehicle ahead, effectively breaking the platoon into subgroups. 
Moreover, whenever the distance becomes smaller, the override mode would be triggered. 
A further limitation is that the dynamic adaptation of the control gain $r$ in~\cite[\secname~6.1]{giordano2019joint} relies on the notion of a desired speed and acceleration, which are not always known a priori.  

As a result, the original controller cannot be directly applied to mixed platoons without modification. 
Different solutions are possible, for example, computing the trigger condition with a \ac{CACC} controller, but the choice of which controller to use is non-trivial. Here, we propose a simple heuristic that is controller-independent and relies only on an acceleration threshold.

\begin{algorithm}[t]
\caption[width=.9\columnwidth]{\gsbl{} follower logic for vehicle $i$.}
\label{alg:autoBraking}
\begin{algorithmic}[1]
\Procedure{init}{}
    \State $s \gets \Cruise{}$
\EndProcedure

\Procedure{onEgoLeaderBeacon}{$v_l$, $u_l$}
    \If{$u_l \geq \SI{0}{\mpsq}$}
        \State $s \gets \Cruise{}$
    \Else
      \State $Cond_1 \gets u_l \leq \delta_a$
      \State $Cond_2 \gets d \leq \SI{4}{\m} \land v_i - v_{i-1} > \SI{0.1}{\mps
      }$ 
      \If{$Cond_1 \lor Cond_2$}
        \State $s \gets \Override{}$
      \EndIf
    \EndIf
    \If{$s = \Override{}$}
        \State $a_{des} = u_l$
        \State $v_{des} = v_l + u_l \cdot \delta_t$
        \State $v_r = v_{des}$
        \State $r = \left|\frac{a_{des}}{v_i - v_{des}}\right|$
    \Else
        \State $v_r = v_l$
        \State $r = \text{default value}$
    \EndIf
\EndProcedure
\end{algorithmic}
\end{algorithm}

\Cref{alg:autoBraking} presents the pseudo-code of this heuristic.
A \gsbl vehicle can be in two states, as in the original paper: \Cruise{} (the initial state) or \Override{}. 
Upon receiving a beacon from its egoLeader, the vehicle extracts the leader's speed $v_l$ and control input $u_l$. 
If $u_l$ falls below a certain threshold $\delta_a$, the vehicle switches to \Override{} and remains there until $u_l$ becomes positive again. 
An alternative condition triggers the same switch when the distance $d$ to the predecessor falls below \SI{4}{\m} while $v_i - v_{i-1} > \SI{0.1}{\mps}$, meaning that the \gsbl vehicle is closing in too fast.  

In override mode, the desired acceleration $a_{des}$ is set equal to $u_l$, and the desired speed $v_{des}$ is extrapolated by projecting the leader's motion $\delta_t$ seconds ahead under constant acceleration. 
These values provide the input needed by the original control law, allowing the reference speed $v_r$ to be set accordingly and the $r$ parameter to be adapted as in~\cite[\eqname~(40)]{giordano2019joint}. 
In \Cruise{} mode, instead, the reference speed is simply $v_l$ and $r$ takes its default value.  

The last step is to define the parameters $\delta_a$ and $\delta_t$, which control when the adaptive behavior in~\cite[\eqname~(40)]{giordano2019joint} is enabled.
Following the error-bounding reasoning in~\cite{giordano2019joint}, we use a look-ahead time of $\delta_t = \SI{1}{\second}$ as a good heuristic for predicting the leader's speed. For the acceleration threshold, we consider alarming a speed reduction over a second approaching \SI{10}{(\kmh)\per\second}, which means an $\approx \SI{-2.7}{\mpsq}$ acceleration. To remain conservative, we set $\delta_a = \SI{-2}{\mpsq}$.

\section{Single Platoon Experiments}
\label{sec:expb}

We implement our evaluation scenarios in \plexe~\cite{segata2022multi-technology}, a framework for simulating \ac{CACC} systems, extending its logic to support platoons where each vehicle may run a different \ac{CACC} algorithm.
This also allows us to combine arbitrary heterogeneous configurations with standard homogeneous platoons.  

In this section, we begin with a set of experiments on isolated platoons to characterize the fundamental behavior of mixed formations.
In \cref{sec:expr}, we then move to more realistic traffic scenarios to evaluate their efficiency and to investigate potential safety issues.  
Each scenario defines a setup in which multiple individual experiments can be executed, collectively building a knowledge base that helps assess the performance we can expect if \ac{CACC} systems are progressively deployed on the roads without a common control framework.

\subsection{Performance of Mixed Platoons}
\label{ss:exp-one}

We consider a single platoon of length \noc with $V_0$ driving autonomously and following a predefined speed pattern. 
The following vehicles implement an arbitrary mix of the \pathc, \ploeg, or \gsbl controllers. 
Our goal is to assess whether a string of \ac{CACC}-enabled vehicles remains stable and how it performs when \cpa, \cpl, and \cgs are arbitrarily combined under proper communication schemes. 
Recall that each vehicle elects as egoLeader the first vehicle ahead running a different control algorithm. 
For instance, in Topology 3 of \cref{fig:topo}, $\cal{P} = \{-, \cpa, \cpl, \cpa, \cpa \}$, so $V_3$ and $V_4$ elect $V_2$ as egoLeader.

We evaluate platoons of $\noc=4$, 8, and 16 vehicles. 
For $\noc=4$ and $\noc=8$, we test all possible combinations of \cpl, \cpa, and \cgs. 
For $\noc=16$, we randomly generate \num{1000} configurations out of the $3^{\noc-1}$ possible combinations. 
The discussion highlights the worst-case results along with selected representative configurations, described in \cref{ss:res-one}.

Two driving patterns are considered. 
In the first, $V_0$ follows a sinusoidal speed profile at \SI{0.1}{\hertz} oscillating between \num{90} and \SI{110}{\kmh}, as commonly used in \ac{CACC} performance studies. 
In the second, $V_0$ drives at a constant \SI{100}{\kmh} before performing an emergency braking maneuver to a full stop with a deceleration of \SI{8}{\mpsq}. 
We refer to these scenarios as ``sinusoidal'' and ``braking,'' respectively.  

Since defining the performance of a mixed platoon is not straightforward, we adopt a comparative approach against well-understood baselines. 
Specifically, we define differential performance by comparing each mixed platoon either against a string of \noc \ac{ACC} vehicles without communication, or against a homogeneous platoon controlled by the same algorithm as the ego vehicle. 
We denote a given mixed platoon as a \textit{configuration} $c$, identified by its sequence of controllers, and use the apex \ca to refer to a string of \noc \ac{ACC}-controlled vehicles. 
We then introduce three performance metrics, described in the following paragraphs and formalized in \cref{eq:deltaA-max,eq:deltaD-max,eq:lcmax}. 
All metrics are defined so that positive values indicate improvements (in comfort, safety, or efficiency), with larger values corresponding to greater improvements.

\paragraph{Comfort}  
For each vehicle in a platoon, we compute the difference between the maximum acceleration experienced in an all-\ac{ACC} configuration and that observed in configuration $c$:  
\begin{equation}
    \Delta_a(c, i) = \max_t |a^{\text{\ca}}_i(t)| - \max_t |a^c_i(t)|
    \label{eq:deltaA-veh}
\end{equation}
A positive $\Delta_a(c, i)$ indicates that configuration $c$ offers smoother driving (lower acceleration/deceleration peaks) compared to standard \ac{ACC}, whereas a negative value indicates harsher driving.

To capture the worst case in the platoon, we take the minimum across all vehicles:  
\begin{equation}
    \Delta_a(c) = \min_{i \in \{1,\ldots,\noc-1\}} \left(\Delta_a(c, i)\right)
    \label{eq:deltaA-max}
\end{equation}
together with the index of the vehicle $V_i$ that produces it.  
This provides a compact measure of travel comfort in configuration $c$, under the assumption that \ac{ACC} represents the baseline of acceptable comfort.

\begin{figure}
\centering
\includegraphics[width=0.95\columnwidth]{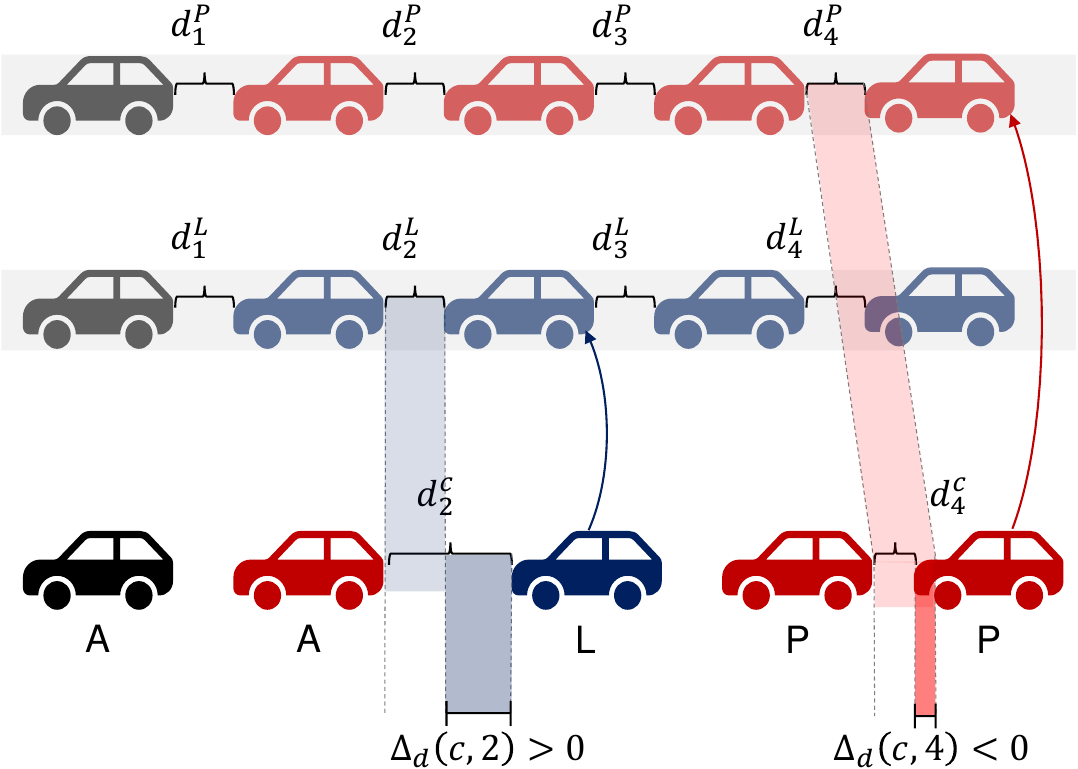}
\caption[width=.9\columnwidth]{Illustration of the $\Delta_d$ metric introduced by \Cref{eq:deltaD}.}
\label{fig:DeltaDist}
\vspace{-5mm}
\end{figure}

\paragraph{Safety}  
To assess safety, we compare the distance $V_{i-1} \leftarrow V_i$ in configuration $c$ with the distance that the same vehicle $V_i$ would maintain in a homogeneous platoon where all vehicles adopt $V_i$'s controller.  
The rationale is that each controller is designed to ensure safety; thus homogeneous platoons can be regarded as safe, both objectively and from the passengers' perspective.  
This comparison therefore provides a relative measure of objective and perceived safety.  

Formally, let $d^c_i(t)$ denote the distance of vehicle $V_i$ from its predecessor at time $t$ in configuration $c$. We define:  
\begin{equation}
    \Delta_d(c, i) =
        \min_t d^c_i(t) -  \min_t d^{\star}_i(t)
    \label{eq:deltaD}
\end{equation}
where $d^{\star}_i(t)$ refer to the distance measured for the same vehicle $V_i$, but in the  experiment where the platoon is homogeneous, i.e., all vehicles use the controller of $V_i$.  
A positive $\Delta_d(c, i)$ indicates that the minimum distance in configuration $c$ is larger than under homogeneous conditions, suggesting that safety is preserved. A large negative value points to potential safety issues.  

To obtain a compact representation for the entire platoon, we consider the minimum across all vehicles (\Cref{eq:deltaD-max}):  
\begin{equation}
    \Delta_d(c) = \min_{i \in \{1,\ldots,\noc-1\}} \left(\Delta_d(c, i)\right)
    \label{eq:deltaD-max}
\end{equation}
\Cref{fig:DeltaDist} illustrates this metric for configuration $c = \{-,\cpa,\cpl,\cpa,\cpa\}$.  
Note that $\Delta_d(c)$ captures the minimum distance observed over the entire experiment, not just on an instantaneous snapshot as as the figure does.

\begin{figure*}[tb]
\centering
\includegraphics[width=\textwidth]{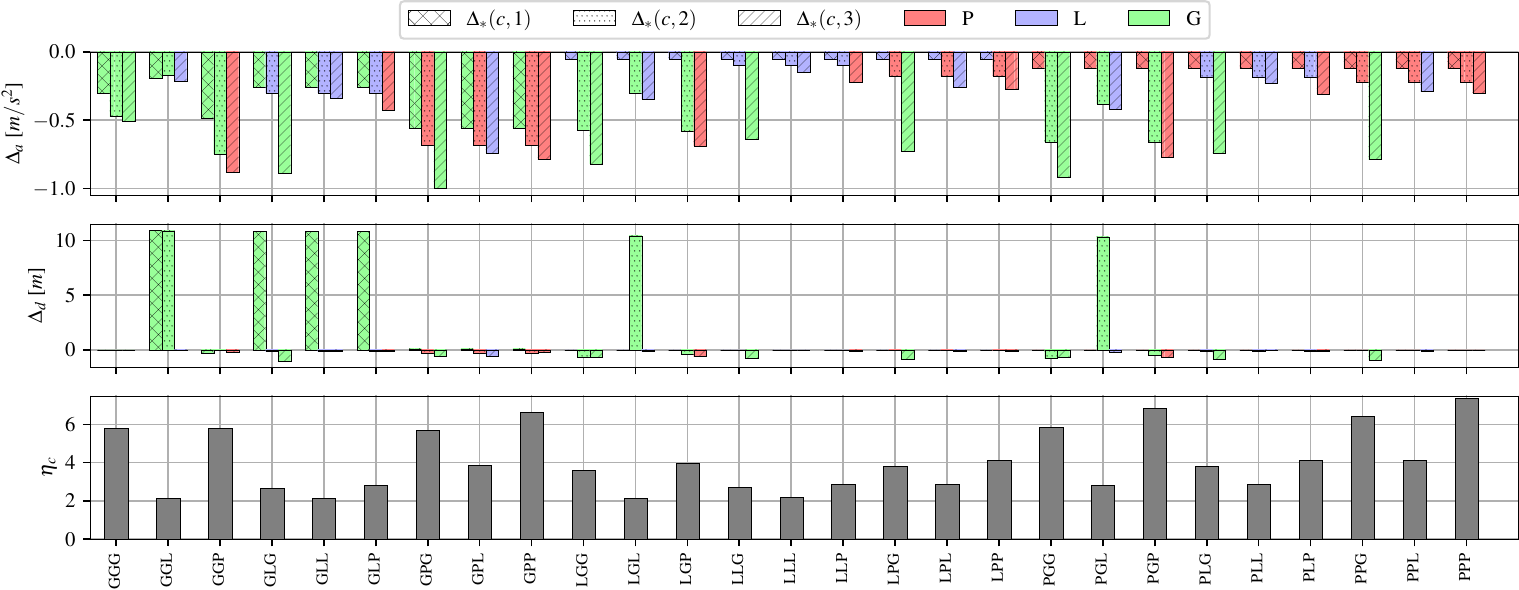} 
\caption[width=.9\textwidth]{$\Delta_a(c,i)$ (top), $\Delta_d(c,i)$ (middle) and $\eta_c$ (bottom) metrics for all possible mixed \ac{CACC} configurations of 4 cars, for the sinusoidal disturbance scenario.}
\label{fig:sinusoidal-exp1-4cars}
\end{figure*}

\paragraph{Efficiency} 
Finally, we consider how effective a configuration $c$ is in increasing the traffic efficiency, i.e., reducing the road occupation. 
Let 
\begin{equation}
    L_{\max}^{c} = \max_t \sum_{i \in \{1,\ldots,\noc-1\}} d^{c}_i(t)
    \label{eq:lcmax}
\end{equation}
be the maximum road occupancy of an arbitrary platoon configuration $c$, homogeneous or not.
The focus is on inter-vehicular distances, while vehicle lengths are not taken into account, as they are identical across configurations and unaffected by the control algorithm.    
We then define  
\begin{equation}
    \eta_c = \frac{L_{\max}^{\ca}}{L_{\max}^{c}}
    \label{eq:length}
\end{equation}
as the efficiency gain relative to a string of \noc \ac{ACC}-controlled vehicles. Larger values of $\eta_c$ indicate better road usage; for example, $\eta_c = 2$ means that the average inter-vehicle distance in configuration $c$ is halved compared to the \ac{ACC} baseline.  

\subsection{Numerical Results}
\label{ss:res-one}

\begin{figure*}[tb]
\centering
\includegraphics[width=\textwidth]{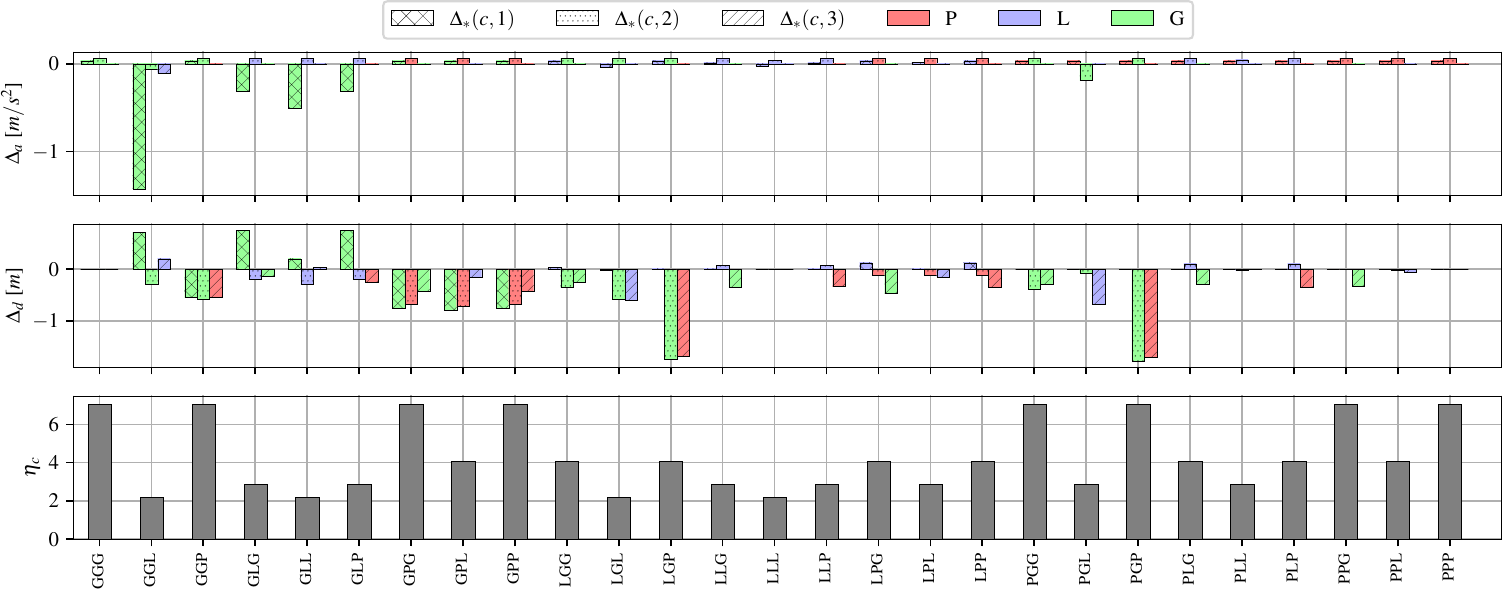} 
\caption[width=.9\textwidth]{$\Delta_a(c,i)$ (top), $\Delta_d(c,i)$  (middle) and $\eta_c$ (bottom) metrics for all possible mixed \ac{CACC} configurations of 4 cars, for the emergency braking scenario.}
\label{fig:braking-exp1-4cars}
\end{figure*}

We begin by analyzing all possible combinations of the three controllers (i.e., \cpa, \cpl, \cgs) in a four-vehicle platoon led by an \ac{ACC} leader.  
\Cref{fig:sinusoidal-exp1-4cars,fig:braking-exp1-4cars} present the three metrics $\Delta_a(c, i)$, $\Delta_d(c, i)$, and $\eta_c$ for the sinusoidal and braking scenarios, respectively.  

In the braking scenario, data collection starts when the first vehicle initiates the emergency maneuver and ends once all cars have slowed below \SI{5}{\kmh}, since values at lower speeds are not representative of the emergency itself. 
In both figures, the plots for $\Delta_d(c, i)$ and $\Delta_a(c, i)$ use colors to indicate the controller type and patterns to indicate the vehicle position. 

\paragraph*{Sinusoidal Scenario}
Focusing on the top plot of \cref{fig:sinusoidal-exp1-4cars}, which reports $\Delta_a(c, i)$, we observe that all controllers yield higher accelerations compared to an \ac{ACC} vehicle part of the reference platoon made of all \ac{ACC}-driven vehicles.
This result is expected, as the smaller inter-vehicle distances require stronger corrections to follow the oscillations of the lead vehicle.  
At the same time, the smaller spacing also results in more stable inter-vehicle distances, which can contribute positively to passenger comfort.  
Mixed platoons show the largest deviations, highlighting that the interaction between different controllers is not trivial to predict.  
Among them, \pathc and \gsbl deviate most from \ac{ACC} accelerations, which is consistent with their control laws: both implement constant spacing rather than mimicking an ``advanced'' \ac{ACC} as \ploeg does.  

The middle plot of the same figure reports $\Delta_d(c, i)$, showing that the distances maintained by vehicles are generally similar to those observed in homogeneous platoons with the same ego-vehicle controller.  
The main exception occurs when a \gsbl-controlled vehicle is followed by one using \ploeg.  
Since \gsbl also relies on information from the following vehicle, the larger inter-vehicle distance enforced by \ploeg effectively pulls back the \gsbl vehicle(s), a behavior consistent with the spring-damper model underlying \gsbl.  
This effect merely increases spacing, reducing efficiency but not compromising safety, and can therefore be regarded as acceptable.

The bottom plot reports the efficiency metric $\eta_c$, which refers to the entire platoon rather than to individual vehicles.  
As expected, cooperative platooning improves efficiency, with gains proportional to the average target spacing of the controllers in the platoon.  
Homogeneous or mixed platoons composed of \pathc or \gsbl vehicles are the most efficient, whereas the presence of \ploeg-controlled vehicles reduces efficiency due to their larger target headway.  
For example, the configuration $\{-,\cpl,\cpl,\cpl\}$ achieves $\eta_c = 2$, very close to the theoretical value expected in a non-perturbed scenario.  
Mixed platoons combining \ploeg and \gsbl generally perform no better than homogeneous \ploeg platoons, while adding \pathc controllers improves efficiency.  
Interestingly, efficiency also depends on the specific ordering of controllers, not only on their presence.  
For instance, $\{-,\cpl,\cgs,\cgs\}$ is more efficient than $\{-,\cgs,\cgs,\cpl\}$, and $\{-,\cpa,\cpl,\cgs\}$ outperforms $\{-,\cpa,\cgs,\cpl\}$.  
These differences can be explained by the pulling effect exerted by \ploeg on \gsbl vehicles. 

\paragraph*{Braking Scenario}
\Cref{fig:braking-exp1-4cars} shows the results for the braking scenario.  
The insights from $\Delta_a(c, i)$ and $\Delta_d(c, i)$ differ from those of the sinusoidal case.  
Accelerations are generally very close to those of \ac{ACC}-controlled vehicles, with the exception of configuration $\{-,\cgs,\cgs,\cpl\}$, where the first \gsbl-controlled vehicle exhibits a significantly larger acceleration.  
This again stems from the larger headway enforced by the trailing \ploeg vehicle, which effectively pulls back the preceding vehicles and alters their dynamics.  
From the safety perspective ($\Delta_d(c, i)$), the overall behavior of mixed platoons is positive.
The only notable exceptions are $\{-,\cpl,\cgs,\cpa\}$ and $\{-,\cpa,\cgs,\cpa\}$, which reduce the minimum distance of the third and fourth vehicles by about one meter, so without hampering safety.  
This behavior appears to result from the interaction between a \gsbl-controlled vehicle and its predecessor using a different controller.  
Efficiency is less relevant in emergency braking scenarios, but for completeness we report $\eta_c$, which remains similar to that of the sinusoidal case.  

Overall, these results suggest that heterogeneous platoons are feasible, but the more the control topology and the goal of the controllers diverge, the more critical their interaction becomes.  
In particular, \gsbl tends to induce behaviors that, in complex scenarios, may lead to uncomfortable or even safety-critical situations.  
 
\begin{table}[tb]
\centering
\begin{scriptsize}
\begin{tabular}{llrrr}
\toprule
   &    $c$    & $\Delta_a(c)$   & $\Delta_d(c)$     & $\eta_c$   \\
\midrule
 S & \texttt{-GP\red{G}} & \textbf{-1.00} $(V_{3})$ & -0.58 $(V_{3})$          & 5.70          \\
 S & \texttt{-GL\red{G}} & -0.89 $(V_{3})$          & \textbf{-0.99} $(V_{3})$ & 2.67          \\
 S & \texttt{-PPP}          & -0.30 $(V_{3})$          & 0.00 $(V_{1})$           & \textbf{7.38} \\
\midrule
 B & \texttt{-\red{G}GL} & \textbf{-1.42} $(V_{1})$ & -0.29 $(V_{2})$          & 2.22          \\
 B & \texttt{-P\red{G}P} & 0.00 $(V_{3})$           & \textbf{-1.78} $(V_{2})$ & 7.06          \\
 B & \texttt{-GGG}          & -0.00 $(V_{3})$          & 0.00 $(V_{1})$           & \textbf{7.07} \\
\midrule
 S & \texttt{-GPGPGP\red{G}} & \textbf{-1.28} $(V_{7})$ & -1.24 $(V_{7})$          & 4.99          \\
 S & \texttt{-GLPPL\red{G}G} & -1.26 $(V_{7})$          & \textbf{-2.28} $(V_{6})$ & 3.26          \\
 S & \texttt{-PPPPPPP}          & -0.54 $(V_{7})$          & 0.00 $(V_{1})$           & \textbf{7.10} \\
\midrule
 B & \texttt{-GGG\red{G}GGL} & \textbf{-1.63} $(V_{4})$ & -1.00 $(V_{6})$          & 2.28          \\
 B & \texttt{-LGL\red{G}GPL} & -0.04 $(V_{1})$          & \textbf{-2.80} $(V_{4})$ & 3.15          \\
 B & \texttt{-GGGPGGG}          & -0.00 $(V_{7})$          & -1.20 $(V_{5})$          & \textbf{7.07} \\
\midrule
 S & \texttt{-GLGLPGPPGPPPGP\red{G}} & \textbf{-1.67} $(V_{15})$ & -1.00 $(V_{15})$          & 3.51          \\
 S & \texttt{-GLPPLLLLLGGLP\red{G}G} & -1.63 $(V_{15})$          & \textbf{-2.50} $(V_{14})$ & 2.76          \\
 S & \texttt{-PPPPPPPPPPPPPPP}          & -0.93 $(V_{15})$          & 0.00 $(V_{1})$            & \textbf{7.09} \\
\midrule
 B & \texttt{-PGPGPGP\red{G}GLPLPPP} & \textbf{-1.73} $(V_{8})$ & -3.48 $(V_{4})$          & 4.47          \\
 B & \texttt{-GPPLL\red{G}LLLGGGGGL} & -0.00 $(V_{15})$         & \textbf{-3.65} $(V_{6})$ & 2.58          \\
 B & \texttt{-GPGPGGGPPPGGPPP}          & 0.00 $(V_{14})$          & -2.91 $(V_{3})$          & \textbf{7.06} \\
\bottomrule
\end{tabular}
\end{scriptsize}
\caption{Worst performance ($\Delta_a(c)$, $\Delta_d(c)$) and best efficiency $\eta_c$ among all configurations $c$ for both the sinusoidal (S) and the braking (B) experiments, with 4, 8 and 16 cars.}
\label{tab:criticalvalues}
\end{table}

\begin{table}[tb]
\centering
\begin{scriptsize}
\begin{tabular}{llrrr}
\toprule
    &    $c$    & $\Delta_a(c)$   & $\Delta_d(c)$      & $\eta_c$   \\
\midrule
 S & \texttt{-PL\red{P}} & \textbf{-0.31} $(V_{3})$ & -0.11 $(V_{3})$          & 4.11          \\
 S & \texttt{-PL\red{P}} & -0.31 $(V_{3})$          & \textbf{-0.11} $(V_{3})$ & 4.11          \\
 S & \texttt{-PPP}          & -0.30 $(V_{3})$          & 0.00 $(V_{1})$           & \textbf{7.38} \\
\midrule
 B & \texttt{-\red{L}LL} & \textbf{-0.02} $(V_{1})$ & 0.00 $(V_{1})$           & 2.22          \\
 B & \texttt{-PL\red{P}} & 0.00 $(V_{3})$           & \textbf{-0.35} $(V_{3})$ & 4.09          \\
 B & \texttt{-PPP}          & 0.00 $(V_{3})$           & 0.00 $(V_{1})$           & \textbf{7.06} \\
\midrule
 S & \texttt{-PLPLPP\red{P}} & \textbf{-0.65} $(V_{7})$ & -0.17 $(V_{4})$          & 4.18          \\
 S & \texttt{-PLPLPP\red{L}} & -0.62 $(V_{7})$          & \textbf{-0.27} $(V_{7})$ & 3.48          \\
 S & \texttt{-PPPPPPP}          & -0.54 $(V_{7})$          & 0.00 $(V_{1})$           & \textbf{7.10} \\
\midrule
 B & \texttt{-LLL\red{L}LLL} & \textbf{-0.07} $(V_{4})$ & 0.00 $(V_{1})$           & 2.22          \\
 B & \texttt{-PPPL\red{P}PL} & 0.00 $(V_{7})$           & \textbf{-0.75} $(V_{5})$ & 4.36          \\
 B & \texttt{-PPPPPPP}          & 0.38 $(V_{7})$           & 0.00 $(V_{1})$           & \textbf{7.06} \\
\midrule
 S & \texttt{-PPPPPPPPPPPPPP\red{P}} & \textbf{-0.93} $(V_{15})$ & 0.00 $(V_{1})$          & 7.09          \\
 S & \texttt{-\red{L}LLLLLLLLLLLLLL} & -0.49 $(V_{15})$          & \textbf{0.00} $(V_{1})$ & 2.19          \\
 S & \texttt{-PPPPPPPPPPPPPPP}          & -0.93 $(V_{15})$          & 0.00 $(V_{1})$          & \textbf{7.09} \\
\midrule
 B & \texttt{-LLLLLLLLLL\red{L}LLLL} & \textbf{-0.07} $(V_{11})$ & 0.00 $(V_{1})$          & 2.22          \\
 B & \texttt{-\red{L}LLLLLLLLLLLLLL} & -0.07 $(V_{11})$          & \textbf{0.00} $(V_{1})$ & 2.22          \\
 B & \texttt{-PPPPPPPPPPPPPPP}          & 0.38 $(V_{15})$           & 0.00 $(V_{1})$          & \textbf{7.02} \\
\bottomrule
\end{tabular}
\end{scriptsize}
\caption{Same results of \cref{tab:criticalvalues}, but mixing only \ploeg and \pathc controllers.}
\label{tab:criticalvaluesNOGIORD}
\end{table}

\Cref{tab:criticalvalues,tab:criticalvaluesNOGIORD} complete the analysis of single-platoon experiments by reporting selected results for platoons of 8 and 16 vehicles.  
In these cases, space limitations prevent us from presenting all simulated configurations ($3^7$ and 1000, respectively).  
We therefore report only the configurations yielding the worst performance in $\Delta_a(c)$ and $\Delta_d(c)$, together with the best efficiency $\eta_c$.  
For comparison, results for 4-vehicle platoons are also included.  

The first column of each table identifies the scenario: Sinusoidal or emergency Braking.  
For each selected configuration, $\Delta_a(c)$, $\Delta_d(c)$, and $\eta_c$ are reported: The vehicle experiencing the worst performance and the corresponding metric value are highlighted in red; for the best $\eta_c$ there is, by definition, no single vehicle to highlight.  
\Cref{tab:criticalvalues} considers all possible mixed configurations, whereas \Cref{tab:criticalvaluesNOGIORD} restricts to mixtures of \ploeg and \pathc only.  
We included this restricted case because \cref{tab:criticalvalues} consistently shows that the worst performance is associated with \gsbl vehicles, making it useful to evaluate how much \gsbl degrades performance compared to platoons without it.  

The analysis of \cref{tab:criticalvalues,tab:criticalvaluesNOGIORD} confirms that mixed platoons can operate correctly, with only marginal deviations in comfort relative to today's \ac{ACC} systems under perturbed cruising (sinusoidal scenario), and maintaining high safety even during emergency braking, where minimum inter-vehicle distances decrease but never approach collision risk.  
Notably, when only \ploeg and \pathc are combined (\cref{tab:criticalvaluesNOGIORD}), performance is nearly identical to \ac{ACC} in terms of comfort ($\Delta_a(c)$), and often superior to homogeneous platoons in terms of safety ($\Delta_d(c)$), as shown by the cases where the baseline is homogeneous.  
In conclusion, the results with platoons of 8 and 16 vehicles reinforce the observation that the most critical situations arise when mixing controllers with significantly different objectives and control topologies.  
Although somewhat expected, the quantitative evidence provided here offers important insights for the design of future real-world systems.  

\section{Large Scale Experiments}
\label{sec:expr}

The positive results of \cref{sec:expb} motivate further investigation into the feasibility of mixing longitudinal cruising controllers on real roads, with particular attention to cases where controllers differ substantially from one another.  
To this end, we conduct a comprehensive set of large-scale experiments to evaluate traffic behavior when homogeneous and mixed platoons coexist with independent vehicles at varying penetration rates.  

To improve experimental efficiency, we model the highway as a \SI{10}{\kilo\meter} ring with \nlane lanes.  
The use of a ring, rather than a linear stretch of road, is supported by both experimental and theoretical studies~\cite{sugiyama2008traffic} as well as simulation-based analyses of shockwaves~\cite{terruzzi2017effects}.  
Our evaluation focuses on safety, stability, and efficiency metrics, in order to determine whether and how the progressive introduction of \ac{CACC} can simultaneously enhance travel experience and road utilization.  
Comfort metrics in such large-scale scenarios are not considered, as they depend too heavily on specific (and random) engagement patterns to provide meaningful insights.

\begin{table}[tb]
\centering
\begin{small}
\begin{tabular}{llr}
        \toprule
        & Parameter & Value\\
        \midrule
        \multirow{9}{*}{\rotatebox{90}{Mobility}} & No.\ of lanes & $M_L=3$ \\
        & Mix of desired                    & \\
        & speeds per lane & \{100, 115, 130\}  $\pm\SI{5}{\kmh}$ \\
        & Vehicles density$^{*}$ & $\tdns=10$, 20, 40, \ldots \\
        &                                    & \ldots, 180 [car/km] \\
        & Self driving models & IDM, ACC \\
        & Controllers          & \cpa, \cpl, \cgs, \textsc{Random Mix}\\
        & Platoons size  $N$  &  \{4, 8, 16\} \\
        & Penetration Rate  $R$ & \{0.25, 0.50, 0.75\} \\ 
        \midrule
        \multirow{14}{*}{\rotatebox{90}{Controllers parameters}} & Powertrain lag & \SI{0.5}{\second}\\
        & EIDM & SUMO defaults \\
        & ACC $\lambda$ & \num{0.1}\\
        & ACC \thead & \SI{1.2}{\second}\\
        & PATH $C_1$ & \num{0.5}\\
        & PATH $\omega_n$ & \num{0.2}\\
        & PATH $\xi$ & \num{1}\\
        & PLOEG \thead & \SI{0.5}{\second}\\
        & PLOEG $k_p$ & \num{0.2}\\
        & PLOEG $k_d$ & \num{0.7}\\
        & GSBL $k$ & \num{0.7} \\
        & GSBL $h$ & \num{0.71} \\
        & GSBL $r$ &  $\sqrt{0.5}$ \\
        & GSBL $r_{min}, r_{max}$ &  $\sqrt{0.5} \leq r \leq 8$ \\
        \bottomrule
        \vspace{-0.5em}\\
        \multicolumn{3}{l}{$^{*}$\tdns refers to the road density, not per-lane density}
\end{tabular}
\end{small}
\caption{Key parameters used in the experiments.}
\label{tab:etwopar}
\end{table}

We consider a scenario where automated vehicles are deployed more rapidly than communication and cooperative driving technologies, as is currently the case.  
Accordingly, the baseline is a highway in which all vehicles implement \ac{ACC} according to \cref{eq:acc}, but lack \ac{V2X} capabilities.  
When platoons are introduced, they operate alongside non-cooperative \ac{ACC} vehicles.  
For completeness, we also include experiments where all vehicles follow the \ac{EIDM} model~\cite{salles2020extending}.  
We do not, however, examine mixed cases of \ac{ACC} and \ac{EIDM}, as their performance would simply be bounded by the two homogeneous cases, so it is not worth reporting.
We consider a  3-lane highway ($\nlane=3$) where non-cooperative \ac{ACC} vehicles drive by default in the rightmost free lane and may overtake in accordance with road rules.  
\ac{CACC}-enabled vehicles are progressively introduced, forming platoons that remain in the lane consistent with their desired speed and do not change it.
For simplicity, platoons are assumed to have constant size, with $\noc=4$, 8, and 16, as in \cref{sec:expb}.  
The free-flow speed of vehicles is randomly distributed around three desired speeds: 100, 115, and \SI{130}{\kmh} with a uniform distribution $\pm\SI{5}{\kmh}$, and this applies also to platoons. 
\Cref{tab:etwopar} summarizes the parameters characterizing all experiments. 

By traffic efficiency we mean road throughput, so we measure it in vehicles per hour.  
Reducing inter-vehicle distance should in principle increase density, but the actual gain as a function of the penetration rate of \ac{CACC} is difficult to predict.  
Moreover, this aspect has not been studied in relation to specific \ac{CACC} algorithms, and the impact of mixed platoons as defined in \cref{sec:modeling} represents a novel contribution.  
Throughput is measured by four equally spaced counting devices placed along the highway ring, labeled N, E, S, and W.  
Each device records the number of passing vehicles in \SI{15}{\second} intervals, producing four time series of throughput values in vehicles per hour:  
$\{ \thrN(t) \}$, $\{ \thrE(t) \}$, $\{ \thrS(t) \}$, and $\{ \thrW(t) \}$.  
From these series we compute averages and other relevant metrics.  
In particular, overall road throughput is obtained as the average of all four sequences, while multiple runs with different random seeds allow the estimation of confidence intervals and levels.  

To evaluate safety and stability, we compute the coefficient of variation of speed, often referred to as the \textit{volatility} of traffic.  
The coefficient of variation $\cfv$ is the ratio between the standard deviation of a series and the absolute value of its average.
Let $\sigma[s]$ and $\text{E}[s]$ be the standard deviation and the average estimators over the series of speed measures $s = \{ s_k \}$.
$\cfv(s)$ is computed as 
$\displaystyle 
\cfv(s) = \frac{\sigma[s]}{|\text{E}[s]|}
$.

Traffic volatility can be defined in different ways, typically depending on the measurements available.  
For example, with a fixed measurement point on the road, the series $\{ s_k \}$ would consist of speed samples from different vehicles at different times as they pass that point.  
In our case, we define volatility as the coefficient of variation of the speed of each individual vehicle during a simulation run.  
Thus, each series $s$ refers to a single vehicle within a single experiment, making the metric a meaningful indicator of both traffic stability and safety.  

A high volatility value implies that a vehicle frequently and abruptly changes speed, a condition that clearly jeopardizes safety.  
Vehicle speed is sampled every \SI{0.5}{\second}.  
Finally, in situations of extreme traffic density, collisions may occur.  
In such cases, the simulation is interrupted, and the corresponding performance points are explicitly marked, with an explanation of the conditions and configurations that led to these outcomes.

\subsection{Numerical Results}
\label{ss:res-two}

We conduct a large number of experiments across different combinations of parameters and penetration rates, but here we report only the subset of results necessary to explain our findings.  
Specifically, we present the free-flow throughput, which serves as an upper bound, the throughput obtained with homogeneous \ac{EIDM} or \ac{ACC}, and the throughput observed when platoons interact with non-cooperative traffic at different penetration rates.  
Platoons may be homogeneous (i.e., all \ploeg, all \pathc, or all \gsbl) or mixed.  

The total number of experiments corresponds to the combination of all variable parameters listed in the \textit{Mobility} section of \cref{tab:etwopar}, namely  
$10 \times (2 + (4 \times 3 \times 3)) = 380$ experiments.  
Here, $10$ is the number of traffic density values, multiplied by two baseline controllers (\ac{EIDM}, \ac{ACC}), plus three homogeneous platooning controllers and one mixed case $4=3+1$, multiplied by $3$ platoon sizes and $3$ penetration rates.  
Each experiment is repeated ten times with different random seeds, resulting in a total of 3800 simulations.  
Since the results are extremely stable, confidence intervals --within $\pm 1$\% of the point estimate at a 95\% confidence level-- are omitted for clarity, as they would only clutter the plots.  

\begin{figure}[tb]
\centering
\includegraphics[width=\columnwidth]{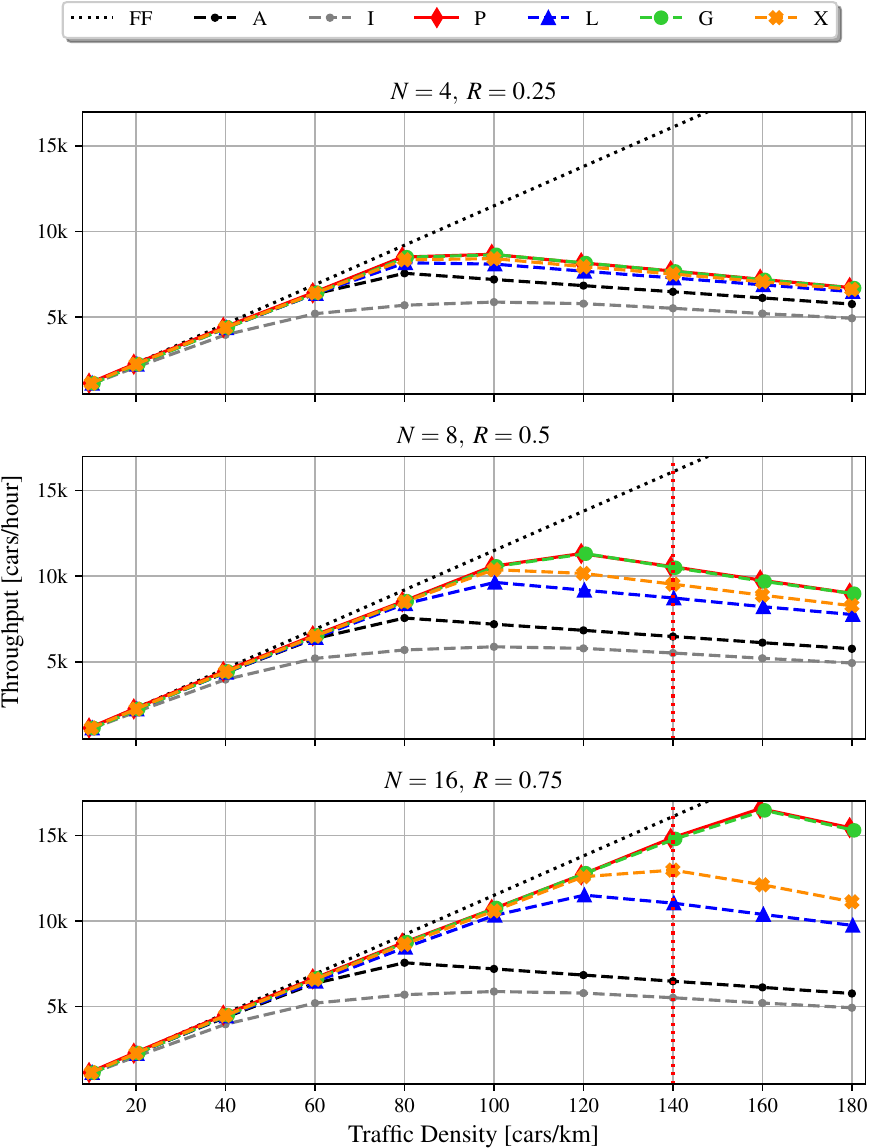}
\caption{Road throughput as a function of the traffic density for three combinations of the platoon size $\noc$ and penetration rate $R$ (180 experiments out of 380). 
The dotted line is the free flow throughput, all the others are the different controllers and self-driving combinations as reported in the legend at the top.} 
\label{fig:exp2-thr-1}
\end{figure}

\Cref{fig:exp2-thr-1} reports road throughput (vehicles per hour) as a function of traffic density, ranging from 10 to 180 vehicles per kilometer.  
For clarity, we show only three representative combinations of platoon size $\noc$ and penetration rate $R$ out of the nine possible, as the others lead to the same conclusions: the top plot refers to $\noc=4, R=0.25$, the middle to $\noc=8, R=0.50$, and the bottom to $\noc=16, R=0.75$.  
In all plots, the black dotted line represents the free-flow throughput, which increases linearly with density.  
The vertical red dotted line marks the density beyond which some simulations terminated due to accidents; these cases are excluded from the reported results.  
Accidents arise from different causes and should not be surprising, as the simulations are realistic and involve extreme traffic conditions.  
Some failures are attributable to imperfect longitudinal control in either \ac{ACC} or \gsbl, while others stem from ``aggressive'' or ``strategically flawed'' lane changes by non-platooning vehicles.  
Notably, all accidents occur in scenarios involving \gsbl controllers, either in isolation or in mixed configurations.  
This observation reinforces the earlier finding that the distinct control topology and objectives of \gsbl make it more challenging to integrate with other cooperative or autonomous\footnote{Recall that all experiment including platoons are done with autonomous \ac{ACC} controlled cars and not with \ac{EIDM} ``human driven'' cars. The investigation of the impact of platoons on traffic with human drivers is outside the scope of this paper, albeit we think it is interesting and deserves further analysis.} driving algorithms.

The first key insight is that the introduction of cooperative vehicles increases road throughput across all traffic densities, with larger gains observed as both the penetration rate of cooperative vehicles and platoon size grow.  
Mixing different controllers within the same platoon (orange line with X markers) does not diminish this efficiency benefit.  
On the contrary, throughput increases roughly in proportion to the efficiency of the controller mix, and mixed platoons consistently outperform homogeneous \ploeg configurations.  
This result is intuitive: \ploeg seeks to mimic standard autonomous \ac{ACC} with modest efficiency gains, whereas \pathc and \gsbl adopt more aggressive strategies to reduce inter-vehicle spacing.

\begin{figure}[tb]
\centering
\includegraphics[width=0.8\columnwidth]{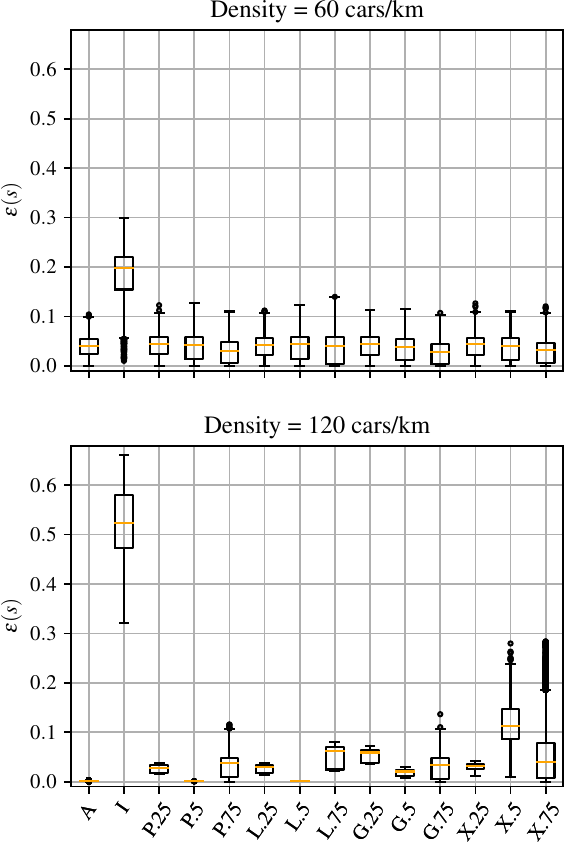}
\caption{Traffic volatility ($\xi(s)$) of all the tested configurations for $\noc=8$, and $\tdns = 60$ (top) and 120 (bottom) cars/km.}
\label{fig:exp2-cvs}
\end{figure}

\Cref{fig:exp2-cvs} shows traffic volatility, estimated by $\xi(s)$, for traffic densities ($\tdns$) of 60 and 120 vehicles/km with $\noc=8$ across all tested configurations.  
Volatility is computed from speed series $s$ of each vehicle, sampled every \SI{0.5}{\second}.  
Each boxplot aggregates $10 \times \tdns \times 10$ values (one per vehicle), corresponding to 10 repetitions of each experiment with $\tdns \times 10$ vehicles on a \SI{10}{\kilo\meter} ring.  
Box edges represent the 25th–75th percentiles, the orange line marks the median, whiskers extend to the last sample within 1.5 interquartile ranges, and outliers appear as isolated circles.  
The first observation is that traffic driven by \ac{EIDM} exhibits substantially higher volatility than any other case, reinforcing the claim that autonomous and cooperative driving will improve travel stability and comfort.  
With \ac{EIDM} and the chosen parameters, shockwaves appear and are observed in the simulations, explaining the elevated volatility. 

It is worth noting that, excluding \ac{EIDM}, most $\xi(s)$ values remain below $\simeq 0.1$, which is very close to the coefficient of variation of the free-flow speed mix used in the simulations, i.e., $\frac{135-95}{\sqrt{3}\cdot 230}\simeq 0.1$.  
This indicates that observed speed variations are mainly due to faster vehicles decelerating before overtaking slower ones.  
Beyond this general trend, the extreme stability of \ac{ACC} and of L.5 (\ploeg at penetration rate 0.5) at $\tdns=120$ is particularly remarkable, as it demonstrates the complete absence of shockwaves or other traffic instabilities.  
Mixed platoons, while performing slightly worse than homogeneous ones, still show acceptable stability, supporting the idea that cooperative driving can be introduced, albeit with some additional care, while preserving the autonomy of manufacturers in designing their own driving algorithms and controllers. 

\section{Modeling Heterogeneous Platoons}
\label{sec:matrix}

The analysis in \cref{sec:expr,sec:expb} leads to very interesting conclusions, obtained only experimentally  due to the lack of a theoretical framework that would allow more general conclusions.
The goal of this section is to propose a method model heterogeneous platoons in a formal setting, to stimulate future research on the topic.

The topology of each of the controllers defined and analyzed so far can be described by the line of a matrix \ccc that represent the logical connectivity of each vehicle with the other vehicles of the platoon.  
\ccc does not describe the control law, but only the use of information related to a given vehicle, either measured with local sensors or distributed via wireless communications. 
For instance, $\ccc_2$ and $\ccc_4$ in \cref{eq:matmod} represent the control topologies 2 and 4 in \cref{fig:topo} respectively, highlighting the fact that the first vehicle (first row) is independent of the others and all other vehicles use also their own data to implement the closed loop control (the main diagonal set to $1$ for rows $2, \ldots, 5$).   

\begin{equation}
\hspace{-4mm}\ccc_2 = 
\begin{pmatrix}
 1 & 0 & 0 & 0 & 0 \\
 1 & 1 & 0 & 0 & 0 \\
 1 & 1 & 1 & 0 & 0 \\
 1 & 0 & 1 & 1 & 0  \\
 1 & 0 & 0 & 1 & 1  \\
\end{pmatrix};~
\ccc_4 = 
\begin{pmatrix}
 1 & 0 & 0 & 0 & 0 \\
 1 & 1 & 1 & 0 & 0 \\
 1 & 1 & 1 & 1 & 0 \\
 0 & 0 & 1 & 1 & 0  \\
 0 & 0 & 1 & 1 & 1  \\
\end{pmatrix} 
\label{eq:matmod}
\end{equation}

The connectivity matrix \ccc makes several features and considerations on mixed platoons evident. 
First of all, if the leading vehicle is not included in the overall platoon control system the first row of \ccc is all zeros but the first element.
This is considered standard today, as it is normally assumed that vehicles are driven based only on the traffic conditions ahead; however, this is not a necessary condition, and in coordinated, cooperative driving using information from the traffic behind may be beneficial. 
Indeed, the \gsbl control system \cite{giordano2019joint} not only uses information from the vehicle behind, but can also use an external speed reference for the platoon leader too, a feature not captured by matrix \ccc, but that can be easily included if needed, for instance with an additional column that models a sort of virtual leader.
If we call this extended matrix \ccc' then the control topology of a 5-vehicle homogeneous platoon with \gsbl controllers would be 
\begin{equation}
\ccc' = 
\begin{pmatrix}
 1 & 1 & 1 & 0 & 0 & 0 \\
 1 & 1 & 1 & 1 & 0 & 0 \\
 1 & 0 & 1 & 1 & 1 & 0 \\
 1 & 0 & 0 & 1 & 1 & 1  \\
 1 & 0 & 0 & 0 & 1 & 1  \\
\end{pmatrix}
\label{eq:cccp}
\end{equation}
while a 6 vehicles heterogeneous platoon $\{\cgs, \cgs, \cpa, \cpa, \cpa, \cpl \}$ topology is described as
\begin{equation}
\ccc' = 
\begin{pmatrix}
 1 & 1 & 1 & 0 & 0 & 0 & 0\\
 1 & 1 & 1 & 1 & 0 & 0 & 0\\
 0 & 0 & 1 & 1 & 0 & 0 & 0\\
 0 & 0 & 1 & 1 & 1 & 0 & 0\\
 0 & 0 & 1 & 0 & 1 & 1 & 0\\
 0 & 0 & 0 & 0 & 0 & 1 & 1 \\
\end{pmatrix}
\label{eq:ccch}
\end{equation}

A second characteristic that emerges from inspecting \ccc and $\ccc'$ is the distinct algebraic structure of controllers whose topology relies solely on information from preceding vehicles, as opposed to those that also incorporate information from vehicles behind.  
In the former case, \ccc is lower triangular, whereas in the latter it is not.  
When some vehicles also track an external reference, $\ccc'$ is not square, so triangularity cannot be defined in the strict sense, although the underlying distinction remains valid.  

It is important to emphasize that the connectivity matrices \ccc and $\ccc'$ do not, by themselves, constitute a control system.  
Nevertheless, they can be easily exchanged among vehicles in a platoon, allowing each vehicle to be aware of the overall platoon composition.  
Once both the platoon topology and the control algorithms of its members are known, the controlled system is fully characterized, and vehicles can adapt their control parameters to enhance safety and performance, similar to how egoLeader selection is handled in \cref{s:mixed-met}.

To the best of our knowledge, heterogeneous control systems have not been analyzed in this manner before.  
Although such an analysis lies beyond the scope of this paper, we believe it represents a promising research direction.  
It could open new perspectives on the progressive deployment of cooperative driving while granting manufacturers greater freedom in the design of cooperative vehicle controllers.

\section{Discussion and Conclusions}
\label{sec:discuss}

The analysis and results presented in this paper are preliminary; nevertheless, they raise important questions and suggest at least three promising research directions.  

The first concerns the composition of messages and the communication capabilities of vehicles.  
Currently, \acp{CAM} do not include information on \ac{CACC} capabilities, yet our results indicate that \ac{ACC} and \ac{CACC} enabled vehicles can cooperate on the road provided they are all \ac{V2X} capable.  
In our study (\cref{sec:expr}), isolated \ac{ACC} vehicles were assumed not to have \ac{V2X}, so the penetration rate of \ac{CACC} referred to already-formed platoons.  
While pre-building platoons is feasible in simulation, it is unrealistic in practice, therefore, future work should explore scenarios where all vehicles are \ac{V2X} equipped and cooperate dynamically to form platoons as preliminarily explored in~\cite{ghirovnc2024}.  
Early experimental works like~\cite{milanes2014cooperative} have (correctly) focused mainly on the safety interaction of small, homogeneous platoons with human driven vehicles; now it is time to look further in the future when \ac{CACC} vehicles will start entering the market.  

The second direction is more theoretical and concerns the fundamental properties of mixed platoons.  
As highlighted in \cref{sec:related}, only a handful of studies address strings of vehicles running different \acp{CACC}, underscoring the need for dedicated theoretical frameworks.  
\Cref{sec:matrix} introduced an initial idea, representing heterogeneous platoons through connectivity matrices, without pursuing a full analysis.  
Developing formal models able to predict the dynamics of mixed control algorithms would be essential to derive performance bounds, guide simulation efforts, and inform the optimal design of \ac{CACC}-enabled vehicles whose coexistence with conventional traffic yields the maximum possible improvement in the road transportation quality.  

The third line of inquiry relates to extending the analysis to additional \acp{CACC} and more advanced cruising controllers in complex traffic scenarios.  
Many-to-many topologies were not considered here, nor were model-predictive consensus algorithms.  
This research path also connects to the study of spontaneous platoon formation, as opposed to centralized optimization, which may be impractical or unnecessary at low penetration rates.  
Ultimately, pursuing this direction could lead to general performance bounds, offering manufacturers concrete guidance on the design limits of cooperative and autonomous cruise control systems.  

\footnotesize
\bibliographystyle{IEEEtranUrldate.bst}
\bibliography{references.bib}
\end{document}